\documentclass[conference,anonymous]{IEEEtran}
\usepackage{cite}
\usepackage{amsmath,amssymb}
\usepackage{graphicx}
\usepackage{booktabs}
\usepackage{tabularx}
\usepackage{array}
\usepackage{url}
\usepackage{algorithm}
\usepackage{algorithmic}
\usepackage[colorlinks=true, linkcolor=black, urlcolor=blue, citecolor=black]{hyperref}
\usepackage{xspace}

\newcommand{\framework}{PRISM-Coach}

\newcommand{\Nusers}{2{,}800\xspace}

\newcommand{\AdhBaseline}{0.35}
\newcommand{\AdhDeployed}{0.68}

\newcommand{\EngIndexPop}{1.35}     
\newcommand{\EngIndexStatic}{0.90}  
\newcommand{\EngIndexAI}{1.33}      

\newcommand{\AdhStatic}{0.48}
\newcommand{\AdhAI}{0.74}
\newcommand{\WLStatic}{3.1}
\newcommand{\WLAI}{5.2}

\begin{document}

\title{PRISM-Coach: Privacy-by-Design Adaptive Group Assignment for Digital Lifestyle Coaching at Scale}


\author{
\IEEEauthorblockN{Nariman Mani}
\IEEEauthorblockA{
\textit{Engineering/R\&D Department, Nutrosal Inc.} \\
Ottawa, ON, Canada \\
nariman@research.nutrosal.com
}
\and
\IEEEauthorblockN{Salma Attaranasl}
\IEEEauthorblockA{
\textit{Engineering/R\&D Department, Nutrosal Inc.} \\
Ottawa, ON, Canada \\
salma@research.nutrosal.com
}
}

\maketitle
\bstctlcite{IEEEexample:BSTcontrol}
\begin{abstract}
Digital lifestyle coaching systems must simultaneously (i) adapt peer support as user behavior and engagement evolve and (ii) prevent personally identifiable information (PII) and sensitive health information from leaking into analytics and AI pipelines. These objectives create a practical tension: personalization requires longitudinal linkability, while privacy engineering requires minimization, separation, and controlled re-identification. We present \framework\ (\emph{Privacy-Restricted Identity Separation and Mapping for Coaching}), a stakeholder-centered architecture and adaptive peer-group assignment method that makes this tension explicit and resolves it through enforceable boundaries. \framework\ separates a single user reality into four bounded views \emph{Identity}, \emph{Operational}, \emph{Learning}, and \emph{Coaching} with distinct access controls and risk profiles. On top of this boundary, we implement vault-based controlled identity restoration, a privacy-constrained contextual bandit that assigns users to eligible peer groups under coach-capacity and stability constraints, and a human-in-the-loop coaching assistant that generates de-identified summaries and draft messages without sending raw PII/PHI to external AI services.

We instantiate \framework\ in a commercially deployed lifestyle coaching platform and evaluate it using three years of telemetry (approximately \Nusers users) and an in-app needs assessment survey. At the population level, daily check-in adherence is \AdhDeployed\ versus \AdhBaseline\ in the pre-deployment baseline, and engagement is  \EngIndexPop$\times$ relative to baseline. In a matched 19-week comparison window (8-week pre, 11-week post), the AI-enabled workflow achieves \AdhAI\ adherence versus \AdhStatic\ under static grouping and yields higher average weight loss (\WLAI\,kg versus \WLStatic\,kg). In surveys, 82\% of users report positive perceived benefit and 92\% report confidence in platform privacy protections after transparency disclosures. These results position \framework\ as a practical blueprint for privacy-by-design learning systems in everyday wellness. To support transparency and reproducibility,  an anonymized, reproducible research artifact with de-identified datasets, analysis code, and reference implementations accompanies this paper.
\end{abstract}


\section{Introduction}
\label{sec:introduction}

Digital lifestyle interventions often underperform in routine use because exposure to the intervention is uneven: users disengage, self-monitoring decays, and accountability weakens over time \cite{eysenbach2005law}. Group-based coaching can counteract this dynamic by creating social reinforcement and accountability, but production deployments expose a deeper systems problem. You must adapt peer-group placement as user behavior and engagement drift, yet you must also protect the sensitive data created by daily logs, posts, and coach messages.

Two theory-to-practice tensions make this problem hard.

\textbf{Tension 1: personalization needs linkability, privacy needs separation.}
Adaptive assignment requires longitudinal features that link behavior over time.
Privacy engineering guidance pushes systems toward minimization, purpose limitation, and controlled re-identification pathways \cite{nist800122,enisa2019pseudonymisation,iso20889}.
A naive ``strip names'' approach fails because behavioral traces and free-text still enable linkage and inference.

\textbf{Tension 2: learning needs exploration, coaching needs stability.}
Contextual bandits can improve personalization by balancing exploration and exploitation \cite{li2010contextual,sutton2018rl}.
Coaching operations impose hard constraints: coach capacity, group-size limits, eligibility policies, and stability requirements that prevent disruptive churn.

This paper argues that you can resolve both tensions by treating privacy and operations as first-class constraints in the learning loop, not as after-the-fact implementation details. We structure the paper around three research questions:

\begin{itemize}
\item \textbf{RQ1:} Can controlled identity separation and mapping enable learning-driven personalization while preventing PII/PHI exposure to AI components and external services?
\item \textbf{RQ2:} Does adaptive peer-group assignment improve adherence and engagement relative to static grouping when you must respect capacity, stability, and eligibility constraints?
\item \textbf{RQ3:} Can human-in-the-loop AI assistance improve coaching throughput and stakeholder experience without weakening privacy guarantees?
\end{itemize}

\textbf{Contributions.}
(1) We define a multi-view privacy boundary model that separates one user reality into four bounded representations \emph{Identity}, \emph{Operational}, \emph{Learning}, and \emph{Coaching} with explicit access controls and auditable controlled restoration.
(2) We formulate adaptive peer-group assignment as a privacy-constrained contextual bandit that filters actions by operational constraints before scoring.
(3) We implement a bounded coaching-assistant workflow that operates only on de-identified inputs and requires coach review for all outgoing messages. 
(4) We evaluate the deployed system using three years of telemetry from approximately \Nusers users and an in-app needs assessment survey, reporting instrumented adherence/engagement alongside stakeholder-reported outcomes.
\textbf{Reproducibility and artifacts.}
To support transparency and reproducibility, we release an anonymized research artifact containing de-identified derived datasets, end-to-end analysis code, and reference implementations of key system components used in this study (Section \ref{sec:artifact}).

\textbf{Reading guide.}
Section~\ref{sec:related} reviews prior work.
Section~\ref{sec:theory_requirements} derives design requirements from the tensions above and from stakeholder evidence.
Sections~\ref{sec:design_overview}--\ref{sec:assistant} present the resulting system design.
Section~\ref{sec:evaluation} evaluates outcomes and privacy properties and answers RQ1--RQ3.

\section{Related Work}
\label{sec:related}

\subsection{Attrition and Measurement in Digital Health}

Attrition and inconsistent engagement are well-documented challenges in digital health interventions. Eysenbach formalized the “law of attrition,” emphasizing that dropout is an intrinsic property of many internet-based interventions and that evaluation must account for usage patterns rather than assuming uniform exposure \cite{eysenbach2005law}. For systems research, this implies that reported outcome improvements require explicit definitions of behavioral exposure, adherence, and engagement.

\subsection{Digital Peer Support and Group-Based Lifestyle Interventions}

Peer support is a recurring mechanism for sustaining behavior change in weight-management and lifestyle programs.
Reviews of peer support groups report that shared community, accountability, and low-friction encouragement can support adherence and maintenance, including in online formats \cite{ufholz2020peer}.
Recent evidence also suggests that digital peer support interventions can improve behavioral, psychosocial, and clinical outcomes, although effect sizes vary by implementation quality and engagement \cite{yeo2025digitalpeer}.

These findings imply two production-grade design constraints.
First, group composition and activity level matter: mismatched cohorts can suppress participation and reduce perceived value.
Second, group processes evolve: treating grouping as a one-time onboarding step causes drift as goals, context, and engagement trajectories change.

\subsection{Adaptive Personalization: Contextual Bandits and Reinforcement Learning}

Personalization in health coaching has increasingly adopted learning-based approaches. Reinforcement learning provides a general framing for sequential decision-making, while contextual bandits focus on action selection using observable context and near-term reward signals \cite{sutton2018rl,li2010contextual}. In health deployments, operational feasibility (capacity, stability, and eligibility) and governance (privacy boundaries and auditability) are first-class constraints.

\subsection{Generative AI as Assistive Infrastructure}

Generative AI can support education, communication, and summarization tasks, yet risks such as hallucination, bias, and data leakage remain significant \cite{dave2023chatgpt,chen2024genai,who2021ethicsai,nistairmf2023}. In \framework, generative AI is bounded: it operates on de-identified inputs, uses structured prompts, and requires coach review before delivery.

\subsection{Privacy-by-Design and Differential Privacy}

Privacy requires minimization, enforceable boundaries, and auditable governance. Differential privacy provides a formal mechanism to bound disclosure risk in aggregated outputs \cite{dwork2006dp}. In \framework, identity separation reduces direct exposure risk, while optional differential privacy mechanisms can be applied to cohort-level analytics exports. 

Our recent work has explored related directions in privacy-preserving personalized coaching, adaptive wellness platforms, and AI-driven adaptive decision-making \cite{mani2025sera_privacy_coaching,mani2025chase_wellness_dt,mani2025aist_adaptive_healing}.

\section{\framework\ Design Overview}
\label{sec:design_overview}

\begin{figure}[t]
\centering
\includegraphics[width=\columnwidth]{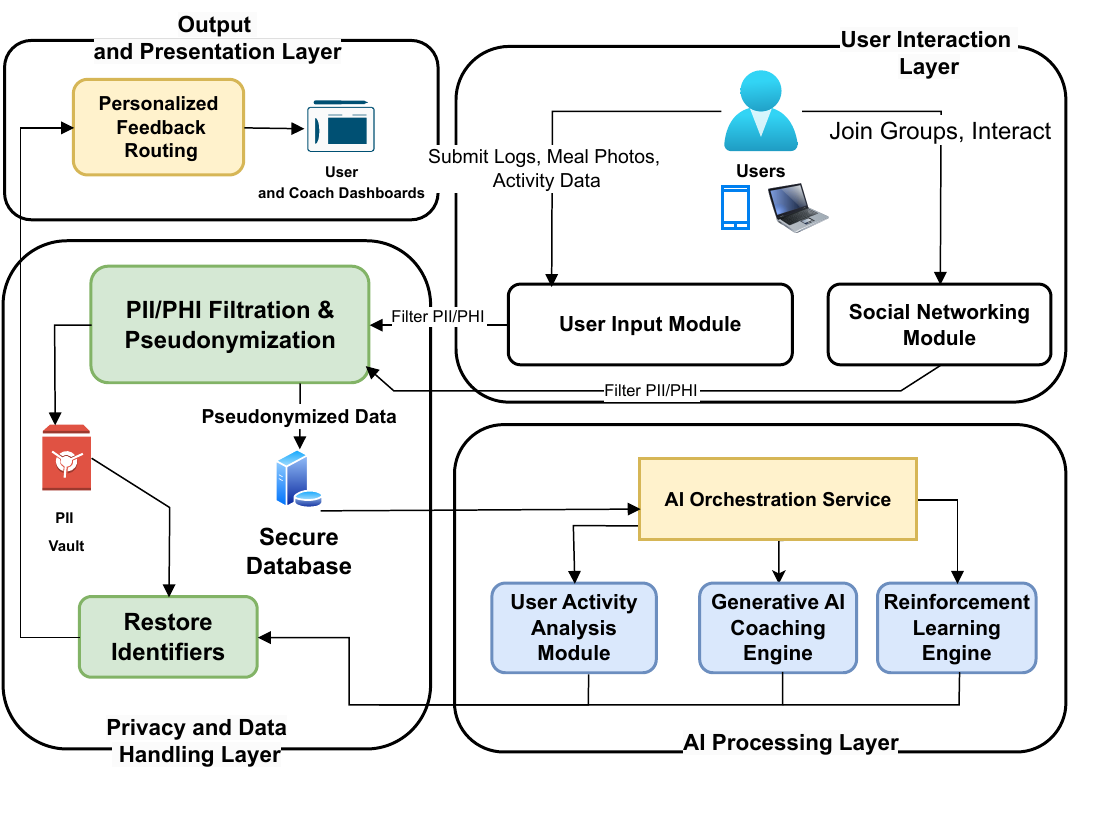}
\caption{System Architecture for Privacy-Preserving AI-Driven Social Networking and Coaching Platform.}
\label{fig:architecture}
\end{figure}

The goal of this section is to demonstrate how the system architecture follows from requirements R1--R4 rather than to document implementation details.
Figure~\ref{fig:architecture} summarizes the end-to-end data flow.
The key claim is structural: user interaction, privacy handling, learning/AI, and stakeholder-facing presentation must be separated so that learning can operate on de-identified representations while operations retain controlled, auditable re-identification.

\subsection{User Interaction Layer}

The User Interaction Layer is the primary entry point for participation and data capture. It contains a \emph{User Input Module} for logging activities, meals (including optional meal photos), and other health-related signals, and a \emph{Social Networking Module} that enables group participation, progress sharing, and coach interaction. This layer directly affects adherence by reducing friction in self-monitoring and by enabling structured peer accountability.

\subsection{Privacy and Data Handling Layer}

The Privacy and Data Handling Layer ensures that sensitive user data (PII/PHI) is filtered and replaced with pseudonymous tokens before it reaches analytics or AI components. It contains: (i) a filtration and pseudonymization service, (ii) a secure operational database holding pseudonymous records, (iii) a protected PII vault holding raw identifiers under stricter controls, and (iv) a controlled restoration service that re-identifies only when required for authorized workflows. Further details are provided in Section~\ref{sec:privacy}.

\subsection{AI Processing Layer}

The AI Processing Layer transforms pseudonymous behavioral data into derived features, risk indicators, and adaptive assignments. It includes feature engineering, activity analysis, and an adaptive assignment engine (this paper: privacy-constrained contextual bandit). It also includes a generative coaching component used in bounded, human-reviewed mode. Details are provided in Sections~\ref{sec:bandit} and \ref{sec:assistant}.

\subsection{Presentation and Output Layer}

The Presentation and Output Layer delivers outputs to stakeholders: users receive dashboards, recommendations, and reminders; coaches receive cohort oversight and AI-generated suggestions for review. This layer is designed for interpretability and accountability, ensuring that system actions remain legible to both users and coaches.

\section{From Theory to Design Requirements}
\label{sec:theory_requirements}

The goal of this section is to justify why the \framework\ design must look the way it does.
We translate the theoretical tensions in Section~\ref{sec:introduction} into concrete system requirements that are testable in deployment.

\textbf{Implication P1 (identity separation must be structural).}
If Learning and AI components can access raw identifiers, accidental leakage becomes a routine failure mode.
Therefore, identity separation must be enforced by the data model and access controls, not by developer convention.

\textbf{Implication P2 (re-identification must be purpose-bound and auditable).}
Operations still require occasional identity restoration (e.g., message delivery, account support).
Therefore, re-identification must flow through a single restoration boundary with RBAC+MFA, immutable audit logs, and rate limits.

\textbf{Implication O1 (constraints must precede learning).}
Production grouping has hard feasibility constraints (capacity, eligibility policies) and soft constraints (stability).
Therefore, the policy must filter infeasible assignments before it scores alternatives.

\textbf{Implication O2 (stability is part of the objective).}
Frequent reassignment can harm trust and group cohesion.
Therefore, the reward and/or scoring rule must penalize churn and enforce minimum dwell time.

\textbf{Design requirements.}
We derive four requirements that drive the remainder of the paper:

\begin{itemize}
\item \textbf{R1 (adaptive personalization):} adapt cohort placement as behavior and engagement drift.
\item \textbf{R2 (structured accountability with harm controls):} create peer accountability while reducing mismatch and inactive cohorts.
\item \textbf{R3 (enforceable privacy governance):} prevent PII/PHI exposure to Learning and AI components by construction; make re-identification controlled and auditable.
\item \textbf{R4 (operational feasibility):} satisfy coach-capacity, group-size, eligibility, and stability constraints in every assignment decision.
\end{itemize}

Table~\ref{tab:rqmap} shows where each research question is analyzed and answered.

\begin{table}[t]
\caption{Research questions mapped to analysis and evidence}
\label{tab:rqmap}
\centering
\small
\begin{tabularx}{\columnwidth}{@{}p{0.9cm}X X@{}}
\toprule
\textbf{RQ} & \textbf{Sections that analyze it} & \textbf{Evidence / where answered} \\
\midrule
RQ1 & Section~\ref{sec:privacy} & Section~\ref{sec:privacy_eval} \\
RQ2 & Section~\ref{sec:bandit} & Section~\ref{sec:comparison} and Tables~\ref{tab:results}--\ref{tab:comparison} \\
RQ3 & Section~\ref{sec:assistant} & Section~\ref{sec:ai_metrics} \\
\bottomrule
\end{tabularx}
\end{table}

\textbf{Stakeholder grounding.}
We validate that these requirements match stakeholder needs through an in-app needs assessment conducted during onboarding and iterative refinement.

\subsection{Needs Assessment Method}
\label{sec:needs_method}

We used two in-app survey instruments.

\textbf{Survey A (needs assessment, onboarding).}
During onboarding, we measured (i) barriers in prior lifestyle programs (multi-select), (ii) baseline attitudes toward motivation/accountability/privacy (Likert), and (iii) optional free-text responses.

\textbf{Survey B (experience and trust, post-exposure).}
After users had at least \textbf{4 weeks} of platform exposure, we administered short Likert items on perceived benefit, peer accountability, and privacy confidence after transparency disclosures.

\textbf{Respondents.}
A total of $N_{\mathrm{surveyA}}=\textbf{1900}$ users completed at least one Survey A item.
For Survey B, item-level denominators vary due to non-response; we compute percentages using the number of non-missing responses for each item (reported alongside each outcome).

\textbf{Computation.}
For multi-select barriers, a respondent may select multiple options.
Percentages in Table~\ref{tab:survey} are computed as:
\begin{equation}
p_b = \frac{n_b}{n_{\mathrm{valid}}}\times 100,
\end{equation}
where $n_b$ is the count selecting barrier $b$ and $n_{\mathrm{valid}}$ is the number of valid responses for that item.

\textbf{Survey governance.}
Participation was voluntary.
We displayed a data-use notice describing how responses inform product improvement and research reporting, and we report only aggregated results in this paper.

\begin{table}[t]
\caption{Needs assessment: barriers in traditional programs (multi-select in-app survey; $N_{\mathrm{survey}}=\textbf{1900}$)}

\label{tab:survey}
\centering
\small
\setlength{\tabcolsep}{3pt}
\begin{tabularx}{\columnwidth}{@{}p{1.65cm}p{0.8cm}X p{1.35cm}@{}}
\toprule
\textbf{Barrier} & \textbf{\%} & \textbf{Interpretation} & \textbf{Contexts} \\
\midrule
Weight regain & 68\% & Initial success followed by loss of adherence; linked to accountability gaps and stress triggers. & Low-carb; calorie counting; IF \\
Monotony / burnout & 34\% & Repetitive plans reduce motivation and increase dropout. & Restrictive diets \\
Insufficient education & 28\% & Limited skill-building for sustainable habits. & Generic programs \\
One-size-fits-all & 52\% & Lack of personalization for constraints and preferences. & Broad apps \\
Limited monitoring & 48\% & Feedback not tailored to individual dynamics and setbacks. & Log-only tools \\
Lack of peer support & 64\% & Weak accountability and social reinforcement. & Individual-only \\
Isolation & 35\% & Reduced resilience during plateaus and stress periods. & Low-touch \\
Low group engagement & 27\% & Group mismatch (goals/activity) prevents meaningful connection. & Large cohorts \\
Privacy concerns & 23\% & Fear of misuse or exposure of sensitive information. & Most platforms \\
\bottomrule
\end{tabularx}
\end{table}

\section{\framework\ Privacy Architecture and Governance}
\label{sec:privacy}

This section specifies the privacy layer with the depth required to justify “privacy-preserving” as an architectural claim. We define a threat model, detail the pseudonymization and vault separation workflow, and describe how the system prevents data leakage to AI services while supporting learning-driven personalization.

\begin{table}[t] \caption{\framework\ views and boundaries} \label{tab:prismviews} \centering \small \begin{tabularx}{\columnwidth}{@{}p{1.55cm}X@{}} \toprule \textbf{View} & \textbf{Content, boundary, primary risk} \\ \midrule Identity & Raw identifiers and sensitive fields stored in a protected vault. Access via RBAC+MFA and audited restoration. Risk: re-identification if vault/keys compromised. \\ \addlinespace Operational & Pseudonymous IDs, logs, group membership, app events. Used by core services. Risk: linkage/inference if leaked. \\ \addlinespace Learning & Feature vectors and aggregates used by adaptive assignment. No raw identity fields. Risk: inference/model inversion (mitigated via minimization and aggregation). \\ \addlinespace Coaching & De-identified summaries and draft messages for coaches. Human review required. Risk: accidental disclosure in free-text (mitigated via redaction and structured templates). \\ \bottomrule \end{tabularx} \end{table}

We use the term \emph{pseudonymisation} to denote replacement of direct identifiers with derived or randomly generated surrogates while retaining controlled linkability under approved workflows.
This distinction from anonymisation is central in privacy engineering guidance and standards, which emphasize risk-based design, minimization, and governance controls for re-identification pathways \cite{nist800122,enisa2019pseudonymisation,iso20889}.

\subsection{Threat Model}

We consider three practical threat classes. \emph{T1} assumes exposure of the Operational view (pseudonymous logs, group membership, and app events) due to database compromise or misconfiguration. \emph{T2} assumes insider misuse in which legitimate access is abused to attempt re-identification. \emph{T3} assumes leakage through AI integration, where sensitive fields might be unintentionally transmitted to external AI services. Table~\ref{tab:threats} summarizes the threat classes and the corresponding primary mitigations enforced by \framework.

\begin{table}[t]
\caption{Threats and primary mitigations in \framework}
\label{tab:threats}
\centering
\small
\begin{tabularx}{\columnwidth}{@{}p{0.55cm}X X@{}}
\toprule
\textbf{T} & \textbf{Attack surface} & \textbf{Primary mitigations} \\
\midrule
T1 & Leakage of pseudonymous operational logs, group membership, and events. &
Identity separation by design; minimization of identifiers; least-privilege access; monitoring of exports and unusual query patterns. \\
\addlinespace
T2 & Insider misuse attempting re-identification beyond approved workflows. &
RBAC+MFA for restoration; purpose-bound access; immutable audit logs; periodic access reviews; rate limiting on restoration endpoints. \\
\addlinespace
T3 & Leakage through AI integration (prompt/response content, external services). &
De-identified Coaching view; automated PII redaction; no vault access from AI components; coach review for all outgoing messages; de-identified prompt/output logging for audits. \\
\bottomrule
\end{tabularx}
\end{table}

\subsection{PII/PHI Filtration and Tokenization Workflow}

Incoming user data may contain sensitive fields $\{f_1, f_2, \ldots, f_n\}$.
The filtration stage detects direct identifiers and other high-risk fields and replaces them with tokens before storage in the Operational database or any downstream use by analytics/AI components.

\textbf{Why tokenization (not plain hashing).}
Many identifiers are low-entropy (e.g., emails, postal codes) and enable dictionary attacks if you rely on unsupervised hashes.
We therefore implement deterministic tokenization using a keyed function whose key remains inside the KMS boundary:
\begin{equation}
T_i = \mathrm{HMAC}_{K_T}\!\left(\mathrm{norm}(f_i)\,\|\,\mathrm{ctx}_i\right),
\label{eq:pseudonym}
\end{equation}
where $\mathrm{norm}(\cdot)$ canonicalizes formatting (e.g., lowercasing emails), $\mathrm{ctx}_i$ binds the token to a field type (e.g., \texttt{email}), and $K_T$ is stored and rotated in a key management service.
The Operational database stores only tokens and pseudonymous activity data required to run workflows.

\textbf{Minimization.}
We do not copy raw PII/PHI into the Operational, Learning, or Coaching views.
When downstream components need stable linkage, they use a user token (Eq.~\ref{eq:hmac}) rather than raw identifiers. We use Eq.~\ref{eq:pseudonym} for field-level tokenization and Eq.~\ref{eq:hmac} for user-level stable linkage.

\subsection{Controlled Identity Mapping and Vault Separation}

To support user-specific delivery (e.g., sending a coach message), identity restoration is performed only through a controlled mapping boundary. The Identity vault stores encrypted PII/PHI data under stricter controls:
\begin{equation}
C_u = \mathrm{Enc}_{K_E}(\mathrm{PII}_u),
\label{eq:encrypt}
\end{equation}
where $\mathrm{Enc}$ is symmetric encryption (e.g., AES-256) and $K_E$ is managed in a key management service boundary. A deterministic token may also be used as the stable vault index, such as:
\begin{equation}
T_u = \mathrm{HMAC}_{K_T}\!\left(\mathrm{norm}(\mathrm{SID}_u)\,\|\,\texttt{user}\right),
\label{eq:hmac}
\end{equation}
where $\mathrm{SID}_u$ is a randomly generated internal subject identifier minted at registration and stored only inside the Identity vault mapping.
Using $\mathrm{SID}_u$ (rather than mutable fields such as email) ensures stable linkage even when user contact details change, while preserving the same KMS-bound key separation described above.

\subsection{Vault Hardening and Governance Controls}

We harden the vault and restoration boundary using operational controls commonly required in production healthcare-adjacent systems:
(1) encryption at rest (AES-256) and strict key governance,
(2) role-based access control (RBAC) with multi-factor authentication (MFA) for restoration operations,
(3) comprehensive audit logs for every restoration request (who, what, when, why),
and (4) rate limiting and anomaly monitoring to detect abnormal restoration patterns.

\subsection{AI Data Minimization and Prompt Safety}

To mitigate \emph{T3}, AI-facing components consume only Learning and Coaching views: pseudonymous identifiers, derived features, and de-identified summaries. The system applies prompt redaction rules excluding names, emails, addresses, and sensitive identifiers. Prompts and outputs are logged in de-identified form for auditability, which enables safety monitoring without creating a second sensitive dataset.

\subsubsection{Structured De-identification for Free-Text}

Free-text fields (posts, messages, captions) represent a dominant leakage channel for accidental identifier disclosure.
We apply a deterministic redaction pipeline prior to AI processing.

\begin{algorithm}[t]
\caption{Prompt De-identification Pipeline}
\label{alg:redaction}
\small
\begin{algorithmic}[1]
\REQUIRE Text $m$, user token $T_u$, redaction rules $\mathcal{R}$
\STATE Detect candidate entities using $\mathcal{R}$ (emails, phone numbers, addresses, dates of birth) and an NER model
\STATE Replace detected entities with typed placeholders (e.g., \texttt{[NAME]}, \texttt{[EMAIL]}, \texttt{[PHONE]})
\STATE Remove or generalize rare identifiers (IDs, order numbers) and geolocation finer than city-level
\STATE Attach only $T_u$ and cohort metadata required by the template
\RETURN De-identified message $m^{\mathrm{deid}}$
\end{algorithmic}
\end{algorithm}

We audit this pipeline by sampling prompts and outputs and computing a leakage rate:
\begin{equation}
\mathrm{LeakRate} = \frac{n_{\mathrm{hits}}}{n_{\mathrm{samples}}},
\end{equation}
where $n_{\mathrm{hits}}$ is the number of samples containing residual identifiers under the same detection rules.

\section{Adaptive Peer-Group Assignment as a Privacy-Constrained Contextual Bandit}
\label{sec:bandit}
The goal of this section is to answer RQ2 by formalizing group assignment as a constrained contextual bandit whose action set is filtered by operational feasibility before any learning-based scoring. We report absolute adherence/engagement for interpretability; the bandit reward uses within-user deltas computed against the pre-assignment baseline window.

Static grouping fails when engagement changes, group activity drifts, or mismatch reduces cohesion. \framework\ models peer-group assignment as a constrained contextual bandit problem in which the system selects an action (group assignment) given a user context and observes short-horizon rewards \cite{li2010contextual}.

\subsection{Feature Engineering and Activity Analytics}

The Learning view is derived from operational events using feature engineering. Numerical features (e.g., average daily calories or weekly activity counts) are normalized:
\begin{equation}
X_{\mathrm{norm}} = \frac{X - X_{\min}}{X_{\max} - X_{\min}}.
\label{eq:norm}
\end{equation}
where $X_{\min}$ and $X_{\max}$ are computed over a rolling cohort-level window.
Categorical inputs (e.g., dietary preference categories) are represented via one-hot encoding. The platform also computes signals for disengagement risk, including missed-check-in streaks and engagement slope over time.

\subsection{Context, Actions, and Reward}

At time $t$, user $u$ has a Learning-view context vector $x_{u,t}$ excluding direct identifiers. The action selects an eligible group $g \in \mathcal{G}_{u,t}$ satisfying capacity and stability constraints. Reward captures short-horizon improvement while penalizing disruptive churn:
\begin{equation}
r_{u,t} = w_A \cdot \Delta \mathrm{Adh}_{u} + w_E \cdot \Delta \mathrm{Eng}_{u} - \lambda \cdot \mathrm{ChurnPenalty}_{u,t}.
\label{eq:reward}
\end{equation}
where $\Delta \mathrm{Adh}_{u}$ and $\Delta \mathrm{Eng}_{u}$ denote changes relative to a fixed pre-assignment baseline window.

\subsection{Operational Constraints}

Capacity constraints enforce group size and coach load limits.
Stability constraints rate-limit reassignment and discourage oscillation.
Eligibility constraints enforce goal compatibility and avoid assigning users into inactive cohorts.

\textbf{Capacity.} Each group $g$ has a maximum size $C_g$ and each coach $c$ has a maximum active load $L_c$:
\begin{equation}
|U_g(t)| \le C_g, \qquad \sum_{g \in \mathcal{G}(c)} |U_g(t)| \le L_c .
\label{eq:capacity}
\end{equation}

\textbf{Stability.} We restrict reassignment using a minimum dwell time $\tau$ and penalize disruptive churn:
\begin{equation}
\mathbb{I}\{t - t^{\mathrm{last}}_{u} < \tau\} \Rightarrow g \in \{g^{\mathrm{current}}_u\}.
\label{eq:stability}
\end{equation}
We define $\mathrm{ChurnPenalty}_{u,t}(g)=1$ when $g \ne g^{\mathrm{current}}_u$ and $t - t^{\mathrm{last}}_{u}<\tau'$, else $0$, with $\tau' \ge \tau$ to discourage oscillation.

\textbf{Eligibility.} We restrict candidate groups using goal match, language/time compatibility, and activity thresholds:
\begin{equation}
g \in \mathcal{G}_{u,t} \Leftrightarrow \mathrm{GoalMatch}(u,g)\wedge \mathrm{Active}(g,t)\wedge \mathrm{PolicyEligible}(u,g).
\label{eq:eligibility}
\end{equation}

Constraint filtering occurs in the Operational view.
The scoring model consumes only Learning-view features.

\subsection{Policy: Constrained LinUCB-Style Selection}

We implement a LinUCB-family selection policy with constraint filtering and a stability penalty \cite{li2010contextual,sutton2018rl}. The algorithm estimates expected reward for each eligible group using a learned linear model and adds an uncertainty bonus to balance exploration and exploitation.

\begin{algorithm}[t]
\caption{Privacy-Constrained Contextual Bandit Group Assignment}
\label{alg:bandit}
\small
\begin{algorithmic}[1]
\REQUIRE Learning-view context $x_{u,t}$, eligible groups $\mathcal{G}_{u,t}$, parameters $\theta$, confidence $\beta$
\STATE Filter $\mathcal{G}_{u,t}$ by capacity, stability, and eligibility constraints
\FOR{each group $g \in \mathcal{G}_{u,t}$}
  \STATE Estimate $\hat{\mu}(x_{u,t}, g) = \theta^\top \phi(x_{u,t}, g)$
  \STATE Compute UCB score $s_g = \hat{\mu}(x_{u,t}, g) + \beta \cdot \sigma(x_{u,t}, g)$
  \STATE Apply stability penalty: $s_g \leftarrow s_g - \lambda \cdot \mathrm{ChurnPenalty}_{u,t}(g)$
\ENDFOR
\STATE Select $g^\star = \arg\max_{g} s_g$
\STATE Assign user to $g^\star$; observe reward $r_{u,t}$ after evaluation window
\STATE Update $\theta$ using $(\phi(x_{u,t}, g^\star), r_{u,t})$
\RETURN $g^\star$
\end{algorithmic}
\end{algorithm}

\section{System Instantiation in an Industry-Deployed Platform}

We instantiate \framework\ in a commercially deployed, social media-inspired lifestyle coaching platform used for health and fitness goals via community engagement and coach oversight.
The system operated in routine production use for three years and served approximately \Nusers users under real coach-capacity and group-stability constraints.
Users join via invitation and are grouped by objectives (e.g., weight loss, healthy eating, maintenance).
Each group is assigned a coach who monitors daily reports on eating and exercise habits.
The platform supports peer interaction through posts, media sharing, group chats, and scheduled one-on-one sessions.
It also supports optional AI-driven meal photo analysis for calorie estimation and daily reminders to keep users aligned with goals. When users enable meal photo analysis, we treat images as sensitive health data: the system strips EXIF metadata, avoids attaching direct identifiers, and processes images via a contracted processor under a no-retention agreement
Raw images are excluded from research exports; only derived nutritional features enter the Learning view.

\subsection{Software and Cloud Architecture}

The production system employs a modern cloud architecture for scalability and availability. The front end is implemented in React.js for a responsive user experience. The backend is implemented in Node.js and supports business logic, real-time communication, and integration with AI services. Data persists in a MySQL database storing pseudonymous user profiles, group assignments, daily reports, and interaction logs. Deployment uses a split-hosting model: front end hosted on Netlify, backend hosted on Heroku, and MySQL hosted on AWS RDS. This separation supports independent scaling of user-facing and backend services.

\subsection{Presentation and Output Capabilities}

Users receive a dynamic dashboard summarizing progress and actionable guidance, real-time notifications (e.g., reminders and milestone messages), and periodic downloadable reports summarizing performance over a specified time frame. Coaches receive a dashboard for cohort oversight, individual progress trends, and AI-generated suggestions routed for review under the Coaching view boundary.

\section{Human-in-the-Loop Coaching Assistant and Safety}
\label{sec:assistant}

The coaching assistant is explicitly assistive. It reduces coach cognitive load and improves responsiveness but does not replace human judgment. The assistant consumes Learning-view features and Coaching-view summaries and generates (i) risk flags indicating likely disengagement and (ii) draft messages designed to encourage low-friction re-engagement.

A conceptual interface for bounded generation is:
\begin{equation}
\mathrm{Advice} = \mathrm{GPT}(\mathrm{DeID\_Summary}, \mathrm{Features}, \mathrm{Template}),
\label{eq:gpt}
\end{equation}
where inputs are de-identified and templated. Coaches review drafts before delivery and may edit, approve, or discard.

This workflow addresses known generative AI risks \cite{dave2023chatgpt,sai2024generative}. The system restricts prompts to de-identified templates, prohibits clinical diagnosis or prescribing language, and logs assistant usage for audit and monitoring. Users can opt out of AI-assisted messaging while remaining in coach-led support.

\section{Deployment, Monitoring, and Learning Health System Loop}

Learning health systems emphasize a closed loop from routine data capture to evidence generation to implementation and measurable outcomes \cite{iom2012bestcare,friedman2015science}.
\framework\ operationalizes this loop through behavioral instrumentation, assignment policy monitoring, and privacy-governance audits. \framework\ instruments daily adherence and weekly engagement actions, monitors policy behavior (reward drift, instability, constraint violations), and monitors governance through periodic review of vault restoration logs and anomalous access patterns. For cohort reporting and research dashboards, \framework\ can apply differential privacy mechanisms to reduce re-identification risk in aggregated outputs \cite{dwork2006dp}.

\section{Evaluation}
\label{sec:evaluation}

\subsection{Ethics, Consent, and Data Governance}
\label{sec:ethics_eval}

This study uses product telemetry and survey responses collected as part of routine platform operation.
We report only aggregated results and de-identified examples.

\textbf{Consent and transparency.}
The platform presents a data-use notice describing operational data processing and research reporting.
Users may opt out of research reporting without losing access to standard coaching features.

\textbf{Data minimization.}
Our analysis uses Learning-view features and pseudonymous Operational-view events.
We do not export Identity-view fields for analytics.

\textbf{Retention and deletion.}
Identity-view data is retained for \textbf{24 months after account closure or last authenticated activity} unless the user requests deletion earlier.
Operational/Learning telemetry used for product quality monitoring and research reporting is retained in \emph{pseudonymous or derived form} for up to \textbf{36 months} to support longitudinal analyses.
User deletion requests propagate across Identity, Operational, Learning, and Coaching views using the linked token identifier.

\subsection{Dataset and Units of Analysis}

We evaluate \framework\ on longitudinal deployment data from approximately \Nusers users over three years. We analyze adherence at the user-day level and engagement at the user-week level. Evidence sources include: needs assessment survey, instrumented platform outcomes, controlled comparison windows between AI-enabled and static assignment, and privacy validation through exposure simulation and governance review. We report engagement indices separately for (i) the full population and (ii) the matched comparison window, because each answers a different estimand.

\subsection{Study Design and Cohort Construction}
\label{sec:design}

We evaluate \framework\ using a mixed-methods design.

\begin{itemize}
\item \textbf{Longitudinal observational analysis} over the full deployment period to quantify adherence and engagement trends.
\item \textbf{Controlled comparison window} using a quasi-experimental cohort design to compare static grouping with AI-enabled adaptive assignment and bounded AI assistance.
\end{itemize}

For the controlled window, we used 1:1 propensity score matching without replacement. Propensity scores were estimated with logistic regression using baseline adherence, baseline engagement score, goal category, and tenure (weeks since onboarding), and we applied nearest-neighbor matching with a caliper of 0.2 standard deviations of the logit propensity score.
We exclude users with fewer than \textbf{21 active days} in the pre-period to reduce noise from immediate dropouts, where an \emph{active day} is a day with at least one instrumented event (e.g., check-in completion, meal/activity log, post, comment, or coach-message interaction).

We define the intervention start date $t_0$ as the deployment date of adaptive assignment for the AI-enabled cohort, with a pre-period of $W_{\mathrm{pre}}$ weeks and post-period of $W_{\mathrm{post}}$ weeks:
\begin{equation}
[t_0-W_{\mathrm{pre}},t_0) \rightarrow \text{baseline}, \qquad [t_0,t_0+W_{\mathrm{post}}) \rightarrow \text{outcome}.
\end{equation}

\subsection{Operational Metric Definitions}

Daily adherence is an instrumented event: $y_{u,d}=1$ if user $u$ completes the required daily check-in on day $d$, else $0$. Cohort adherence over a period $D$ is:
\begin{equation}
\mathrm{Adh} = \frac{1}{|U|}\sum_{u \in U}\left(\frac{1}{|D|}\sum_{d \in D} y_{u,d}\right).
\label{eq:adherence}
\end{equation}

\textbf{Weekly engagement score.} We define a weekly engagement score
\begin{equation}
S_{u,w}=\sum_{k=1}^{K}\alpha_k \cdot \tilde{c}_{u,w}^{(k)},
\end{equation}
where $\tilde{c}_{u,w}^{(k)}$ denotes a bounded, non-negative normalized count so that $\mathbb{E}[S_{u,w}]>0$ and the post/pre ratio in Eq.~\ref{eq:engindex} is well-defined. 
Specifically, for each action type $k$ we winsorize raw counts $c_{u,w}^{(k)}$ to the $[5,95]$ percentile range estimated on the \emph{pre-period} distribution, then scale to $[0,1]$:
\[
\tilde{c}_{u,w}^{(k)}=\frac{\min(\max(c_{u,w}^{(k)},P_5^{(k)}),P_{95}^{(k)})-P_5^{(k)}}{P_{95}^{(k)}-P_5^{(k)}+\epsilon},
\]
with $\epsilon=10^{-6}$ for numerical stability and $\sum_k \alpha_k=1$.

\textbf{Engagement index (relative).} To report interpretable changes, we define an engagement index relative to a reference period or cohort:
\begin{equation}
\mathrm{EngIndex}=\frac{\mathbb{E}[S_{u,w}\mid \text{post}]}{\mathbb{E}[S_{u,w}\mid \text{pre}]},
\label{eq:engindex}
\end{equation}
so $\mathrm{EngIndex}=1.35$ indicates a 35\% increase relative to baseline. We compute EngIndex using cohort-level means of $S_{u,w}$ over the relevant pre/post windows. Because $S_{u,w}$ is defined as non-negative (above), EngIndex values below 1.0 indicate decreases, and values above 1.0 indicate increases. EngIndex is computed within each analysis window as the post/pre ratio (Eq.~\ref{eq:engindex}).  Population-level and matched-window EngIndex values need not coincide because they use different cohorts and denominators.

User-reported outcomes are collected using in-app Likert (5-point) items; we report the fraction of responses at 4--5 (“positive”).
\subsection{Statistical Analysis}
\label{sec:stats}

We compute 95\% confidence intervals for all primary telemetry outcomes. Due to space, the main paper reports point estimates, while the artifact (Section~\ref{sec:artifact}) contains the full CI tables and analysis code.

\textbf{Adherence.} Because $y_{u,d}$ is binary, we estimate adherence differences using difference-in-means with cluster-robust standard errors at the user level.
As a sensitivity analysis, we fit a mixed-effects logistic regression:
\begin{equation}
\log\frac{\Pr(y_{u,d}=1)}{1-\Pr(y_{u,d}=1)} = \beta_0 + \beta_1 \cdot \mathbb{I}\{\mathrm{AI\_enabled}\} + b_u + \gamma^\top z_{u,d},
\end{equation}
where $b_u$ is a user random intercept and $z_{u,d}$ includes seasonality controls. We included fixed effects for calendar week-of-year and calendar year, and adjusted for tenure (weeks since onboarding).

\textbf{Engagement.} We compare weekly engagement using a Mann--Whitney U test and a linear mixed model as a robustness check.

\textbf{Multiple comparisons.} For stratified analyses, we control false discovery using Benjamini--Hochberg at $q=0.05$.

\subsection{User-Reported Outcomes and Trust}
\label{sec:survey_analysis}

We collected in-app Likert (5-point) items on perceived benefit, personalization quality, and privacy trust.
We report the fraction of responses at 4--5 (``positive'').

In Survey B (post-exposure), 82\% reported positive perceived benefit (4--5 on a 5-point Likert item; $n=900$).
72\% reported that peer accountability improved motivation ($n=820$).
After PRISM transparency disclosures, 92\% reported confidence in the platform's privacy protections ($n=700$).

We report these perceptions separately from telemetry-derived adherence and engagement metrics to avoid conflating self-report with instrumented behavior.

\subsection{Controlled Comparison Window (Static vs.\ AI-Enabled)}
\label{sec:comparison}

We evaluate RQ2/RQ3 using a quasi-experimental matched cohort design over a 19-week matched comparison window (8-week pre, 11-week post).
We compare an AI-enabled condition (adaptive assignment + bounded assistant workflow) against a static-assignment condition.

\textbf{Cohorts and matching.}
AI-enabled cohort: $N_{\mathrm{AI}}=\textbf{1400}$ users.
Static cohort: $N_{\mathrm{static}}=\textbf{1400}$ users.
We match users using pre-period behavior over $W_{\mathrm{pre}}=\textbf{8}$ weeks (baseline adherence, baseline engagement score, goal category, and tenure).
We exclude users with fewer than \textbf{21} active days in the pre-period to reduce noise from immediate dropouts.

\textbf{Outcomes.}
Over $W_{\mathrm{post}}=\textbf{11}$ weeks post-intervention, the AI-enabled cohort shows higher daily check-in adherence (\AdhAI\ vs.\ \AdhStatic) and higher relative engagement (EngIndex \EngIndexAI\ vs.\ \EngIndexStatic). Unless otherwise noted, Table~\ref{tab:comparison} reports \emph{post-period} means computed over the 11-week outcome window. Pre-period balance checks and pre/post change estimates are provided in the artifact.
Among users with recorded weights in both periods ($N_{\mathrm{wt}}=\textbf{1100}$), average weight loss is \WLAI\,kg in the AI-enabled cohort versus \WLStatic\,kg in the static cohort.

\textbf{Why baseline adherence differs from Table~\ref{tab:results}.}
The population-level baseline (\AdhBaseline) includes early dropouts.
The matched window excludes early dropouts by construction, so baseline adherence in this restricted sample is higher.

\subsection{Case Study Evidence}

Qualitative case studies complement instrumented outcomes by illustrating mechanisms. For example, a user with thyroid issues and stress-related eating reported difficulty maintaining restrictive diets. Under the deployed workflow, the user lost 6\,kg over three months and reported fewer stress-triggered eating episodes, consistent with timely coach outreach supported by risk flags and low-friction re-engagement prompts.

\subsection{AI Performance Metrics}
\label{sec:ai_metrics}

We evaluate AI components with task-specific operational metrics. For coaching suggestions, we compute precision as the fraction of AI suggestions judged actionable by coaches and recall as the fraction of recurring coach-identified issues that the assistant successfully flagged. Under this evaluation, coaching suggestions achieved 88\% precision and 84\% recall. For group assignment quality, we compute accuracy as the fraction of assignments that met goal-alignment and activity-compatibility criteria and sustained positive engagement; accuracy reached 91\%. For activity insights, 78\% of users received at least one actionable insight associated with subsequent positive behavior change. Metrics are computed over coach-labeled review logs from the same matched window, with actionable labels recorded at the time of message approval.

\begin{table}[t]
\caption{AI component performance (operational evaluation)}
\label{tab:aimetrics}
\centering
\small
\begin{tabular}{@{}lcc@{}}
\toprule
\textbf{Component} & \textbf{Metric} & \textbf{Value} \\
\midrule
Coaching assistant & Precision / Recall & 0.88 / 0.84 \\
Group assignment & Accuracy & 0.91 \\
Activity insights & Actionable rate & 0.78 \\
\bottomrule
\end{tabular}
\end{table}

\subsection{Privacy and Security Validation}
\label{sec:privacy_eval}

We validate privacy claims against the threat model. Under exposure simulation (\emph{T1}), operational data remains non-actionable for re-identification without access to the controlled mapping boundary and cryptographic keys (Eqs.~\ref{eq:pseudonym}--\ref{eq:hmac}). Penetration testing of the PII vault detected no unauthorized access, and data exposure simulations confirmed that pseudonymized tokens could not be reversed without vault authorization and keys. After transparency disclosures explaining PRISM boundaries, 92\% of surveyed users reported confidence in platform privacy protections. 

\textbf{De-identification leakage audit.}
We sampled $n_{\mathrm{samples}}=\textbf{1200}$ assistant prompts/outputs uniformly over the matched window and re-ran the same detection rules used in Algorithm~\ref{alg:redaction}.
We observed $\mathrm{LeakRate}=\textbf{0.17 \%}$ ( $\textbf{2}/\textbf{1200}$ residual identifier hits ). Both residual hits were due to user self-disclosure in free-text signatures; no emails, phone numbers, or addresses were observed.

\textbf{Population-level pre/post definition.}
For Table~\ref{tab:results}, ``Pre'' denotes the production period immediately before \framework\ deployment (legacy identity-handling and static grouping), and ``Post'' denotes the period after full \framework\ rollout. 
We compute each metric over all eligible user-days/weeks in the corresponding period; the artifact reports the exact calendar ranges and denominators.

\begin{table}[t]
\caption{Population-level outcomes before vs.\ after \framework\ deployment (telemetry and surveys)}
\label{tab:results}
\centering
\small
\renewcommand{\arraystretch}{1.1}
\begin{tabular}{@{}
  p{0.52\columnwidth}@{}
  p{0.16\columnwidth}@{}
  p{0.16\columnwidth}@{}
  p{0.16\columnwidth}
@{}}
\toprule
\textbf{Metric} & \textbf{Pre} & \textbf{Post} & \textbf{Change} \\
\midrule
Daily check-in adherence (telemetry; Eq.\ \ref{eq:adherence}) & \AdhBaseline & \AdhDeployed & +0.33 \\
Engagement index (telemetry; Eq.\ \ref{eq:engindex}) & 1.00 & \EngIndexPop & +35\% \\
Perceived benefit (survey; positive) & -- & 82\% & -- \\
Privacy confidence (survey; positive) & -- & 92\% & -- \\
\bottomrule
\end{tabular}
\end{table}

\begin{table}[t]
\caption{Matched 19-week comparison window: static grouping vs.\ AI-enabled workflow. For EngIndex, we also show the implied relative change ($(\mathrm{EngIndex}-1)\times 100$) and report the net change in percentage points (pp).}
\label{tab:comparison}
\centering
\small
\renewcommand{\arraystretch}{1.1}
\begin{tabular}{@{}
  p{0.52\columnwidth}@{}
  p{0.16\columnwidth}@{}
  p{0.16\columnwidth}@{}
  p{0.16\columnwidth}
@{}}
\toprule
\textbf{Metric} & \textbf{Static} & \textbf{AI-enabled} & \textbf{Difference} \\
\midrule
Daily check-in adherence (telemetry) & \AdhStatic & \AdhAI & +0.26 \\
Average weight loss (kg) & \WLStatic & \WLAI & +2.1 \\
Engagement index (post/pre; Eq.\ \ref{eq:engindex}) & \EngIndexStatic\ ($-10\%$) & \EngIndexAI\ ($+33\%$) & $+43$ pp \\
\bottomrule
\end{tabular}
\end{table}

\section{Future Work}

Future work will evaluate generalization beyond lifestyle coaching to chronic disease management (e.g., diabetes, hypertension). This includes integrating condition-specific constraints and coaching protocols, adapting group formation to disease-specific support needs, and expanding fairness audits to ensure that adaptive assignment does not systematically disadvantage new or low-activity users.

\section{Ethical, Social, and Stakeholder Implications}

Socio-technical wellness platforms warrant explicit discussion of autonomy, group dynamics, fairness, and trust. \framework\ mitigates over-reliance by keeping AI assistive and requiring coach review for outgoing messages. It mitigates group harm through stability constraints, eligibility thresholds that reduce inactive cohorts, and moderation pathways. It addresses fairness by auditing outcomes across goal categories and engagement strata and by exposing coach-facing rationales for assignments. Finally, it treats privacy as trust infrastructure: user-facing transparency disclosures and auditable restoration boundaries are necessary for adoption.

\section{Discussion and Limitations}

\framework\ advances beyond a descriptive system narrative by making three elements explicit and testable: (i) privacy as a boundary model with controlled mapping and governance, (ii) adaptive assignment as a constrained contextual bandit with measurable reward signals, and (iii) stakeholder-centered integration of assistive AI with human oversight.

Limitations include deployment confounders (seasonality, evolving coaching practice), potential self-report bias in user-reported outcomes and weight, and residual inference risk from behavioral traces even under pseudonymization. These limitations motivate stronger quasi-experimental designs, expanded fairness audits, and differential privacy for cohort analytics exports. Sustained adherence is also economically relevant, as obesity is associated with substantially higher medical expenditures, especially for obesity-linked conditions such as diabetes \cite{cawley2012obesity}.

\subsection{Answers to Research Questions}
\label{sec:rq_answers}

\textbf{RQ1 (privacy boundaries enable learning without PII/PHI exposure to AI).}
RQ1 is answered by the PRISM multi-view boundary design (Identity, Operational, Learning, Coaching) and the enforced rule that AI components consume only Learning-view features and Coaching-view de-identified summaries (Section~\ref{sec:privacy}).
The threat model and mitigations specify how leakage via operational compromise, insider misuse, and AI integration is constrained through vault separation, controlled restoration, RBAC+MFA, and prompt de-identification (Table~\ref{tab:threats}; Algorithm~\ref{alg:redaction}).
We evaluate these controls through exposure simulation and user-reported privacy confidence (Section~\ref{sec:privacy_eval}; Table~\ref{tab:results}).

\textbf{RQ2 (adaptive assignment improves adherence/engagement under real constraints).}
RQ2 is answered by the constrained contextual bandit formulation and policy that filters candidate groups under capacity, stability, and eligibility constraints (Section~\ref{sec:bandit}; Eq.~\ref{eq:capacity}--\ref{eq:eligibility}; Algorithm~\ref{alg:bandit}).
Evidence appears in telemetry outcomes summarized before vs.\ after deployment (Table~\ref{tab:results}) and in a matched comparison window against static grouping (Section~\ref{sec:comparison}; Table~\ref{tab:comparison}).
We treat these findings as associations under a quasi-experimental design and acknowledge remaining confounders in Section~\ref{sec:design}. 

\textbf{RQ3 (human-in-the-loop AI improves coaching throughput/experience without weakening privacy).}
RQ3 is answered by the bounded assistant workflow: templated de-identified inputs, prohibition of diagnosis/prescribing language, mandatory coach review, and audit logging (Section~\ref{sec:assistant}).
Evidence includes coach-judged utility metrics for suggestions (Section~\ref{sec:ai_metrics}; Table~\ref{tab:aimetrics}) alongside the privacy constraint that the assistant never accesses Identity-view data (Section~\ref{sec:privacy}).

\section{Conclusion}

We presented \framework, a stakeholder-centered systems architecture and adaptive peer-group assignment method for digital lifestyle coaching at scale. PRISM enforces four bounded views Identity, Operational, Learning, and Coaching that support learning-driven personalization while minimizing PII/PHI exposure to AI components. In a three-year real-world evaluation with approximately \Nusers users, \framework\ improved adherence, engagement, perceived benefit, and privacy confidence while maintaining enforceable governance boundaries. The results support \framework\ as a practical blueprint for privacy-preserving, stakeholder-centered learning health systems in everyday wellness.

\section{Artifact Availability and Reproducibility}

We provide access to an anonymized artifact package accompanying this paper at the following archive:

\label{sec:artifact}
\noindent
Artifacts: \url{https://doi.org/10.5281/zenodo.18518472}

The archive contains de-identified datasets, analysis scripts, and reference implementations to reproduce the paper’s results under the stated privacy boundaries, while excluding raw PII/PHI and production details.

\bibliographystyle{IEEEtran}
\bibliography{references}

\end{document}